\documentstyle[11pt]{article}

\def\qvec{{\bf q}}
\def\pvec{{\bf p}}
\def\calW{{\cal W}}
\def\calR{{\cal R}}
\def\calA{{\cal A}}
\def\oeps{{\bar\epsilon}}
\def\osig{{\bar\sigma}}
\def\erf{\mbox{erf}}

\title{An Improved Acceptance Procedure for the \\[6pt]
             Hybrid Monte Carlo Algorithm}

\author{{\Large \rule{0pt}{26pt}Radford M. Neal} \\[12pt]
         Department of Computer Science \\
         University of Toronto \\
         10 King's College Road \\
         Toronto, Canada \ \ M5S 1A4 \\[6pt]
	 e-mail: radford@cs.toronto.edu}

\date{27 July 1992}

\begin{document}

\maketitle

\vspace*{0.3in}

\begin{center} {\large \bf Abstract} \end{center}

\noindent The probability of accepting a candidate move in the hybrid Monte
Carlo algorithm can be increased by considering a transition to be
between windows of several states at the beginning and end of the
trajectory, with a state within the selected window being chosen
according to the Boltzmann probabilities. The detailed balance
condition used to justify the algorithm still holds with this
procedure, provided the start state is randomly positioned within its
window. The new procedure is shown empirically to significantly
improve performance for a test system of uncoupled oscillators.

\vspace{0.3in}
\begin{center} (Figures 2, 3, and 4 are not present in this version.)
\end{center}

\pagebreak

\section{Introduction}

The hybrid Monte Carlo algorithm of Duane, Kennedy, Pendleton, and
Roweth \cite{duane} is a method of sampling from complex distributions, such
as those encountered in statistical physics, that combines the
advantages of dynamical methods \cite{anderson} with those of the
Metropolis Monte Carlo method \cite{metropolis}. These and related
methods are reviewed by Toussaint \cite{toussaint}.

The problem is to simulate a system parameterized by a vector,
$\qvec$, of dimension $N$, with a differentiable potential energy
function, $E(\qvec)$. This energy function induces a Boltzmann
distribution over $\qvec$, for which the probability density is
\begin{eqnarray}
  p(\qvec) & = & Z_E^{-1} \exp(-E(\qvec))
\label{distq}\end{eqnarray}
where $Z_E = \int_{R^N} \exp(-E(\qvec))\, d\qvec$. (A temperature of
one is assumed throughout, for simplicity.)

The aim of the simulation is to estimate the expectation of some
function, $h(\qvec)$:%
\begin{eqnarray}
  \langle h \rangle & = &
     \int_{R^N} h(\qvec)\, p(\qvec)\,d\qvec
  \,\ \approx \,\ {1 \over n} \sum_{i=0}^{n-1} h(\qvec_i)
\label{exp}\end{eqnarray}
where $\qvec_0,\qvec_1,\ldots,\qvec_{n-1}$ are obtained from the
simulation, and are each distributed according to the Boltzmann
distribution for $\qvec$ (but are not, in general, independent).

I will first describe the dynamical approach to solving this problem,
which suffers from systematic error in the sampling, and then describe
the hybrid Monte Carlo method, which eliminates this error by
accepting only some of the dynamical moves. Next, I present and
justify a generalization of this algorithm in which moves are made
between windows of states at the beginning and end of a trajectory,
rather than between single states. Situations in which this algorithm
will have a higher acceptance probability than the standard algorithm
are then illustrated, and an empirical comparison is given for systems
of uncoupled oscillators. Finally, I discuss two variations on the
algorithm that may be useful in some circumstances.

\section{The dynamical method}

In the dynamical simulation method, we introduce a momentum vector, $\pvec$,
which, like $\qvec$, has dimension $N$, and a Hamiltonian function,
$H(\qvec,\pvec)$, that incorporates both potential and kinetic energy:
\begin{eqnarray}
  H(\qvec,\pvec) & = & E(\qvec)\ +\ {\textstyle{1\over2}}\,|\pvec|^2
\label{ham}\end{eqnarray}

We then seek to sample from the Boltzmann distribution that $H$
induces on the phase space, $(\qvec,\pvec)$, for which the density
is
\begin{eqnarray}
  p(\qvec,\pvec) & = & Z_H^{-1} \exp(-H(\qvec,\pvec))
                 \,\ =\,\ p(\qvec)\, p(\pvec)
\label{distqp}\end{eqnarray}
where $Z_H = \int_{R^N} \int_{R^N} \exp(-H(\qvec,\pvec))\,d\qvec\,d\pvec$, and
\begin{eqnarray}
  p(\pvec) & = & (2\pi)^{-N/2} \exp(-{\textstyle{1\over2}}|\pvec|^2)
\label{distp}\end{eqnarray}

This sample is generated by simulating an ergodic Markov chain that
has the Boltzmann distribution for $(\qvec,\pvec)$ as its stationary
distribution. Values of $\qvec$ from successive states of this Markov
chain, whose marginal distribution is that of equation (\ref{distq}),
are used to estimate $\langle h \rangle$ using equation (\ref{exp}).

This Markov chain operates by alternating dynamical transitions with
stochastic transitions.

The dynamical transitions consist of simulating the system for some
predefined period in a fictitious time, $\tau$, using Hamilton's equations:
\begin{eqnarray}
  {d\qvec \over d\tau} & = & + {\partial H \over \partial \pvec}
                     \,\ = \,\ \pvec
\label{dyns}\\[3pt]
  {d\pvec \over d\tau} & = & - {\partial H \over \partial \qvec}
                     \,\ = \,\ - \nabla E(\qvec)
\label{dyne}\end{eqnarray}
These equations leave $H$ invariant. Furthermore, the volume of a
region of phase space remains constant as it evolves according to this
dynamics (Liouville's theorem). In consequence, the dynamical
transitions sample regions of constant $H$ without bias.

The stochastic transitions allow regions with different values of $H$
to be explored. They consist of replacing $\pvec$ with a value picked
from its Boltzmann distribution (equation (\ref{distp})). Such
transitions clearly leave the Boltzmann distribution of
$(\qvec,\pvec)$ with respect to $H$ invariant. The presence of these
stochastic transitions will also usually be enough to ensure that the
Markov chain is ergodic --- i.e.\ that the system can move to any point
in phase space.

It is generally desirable that the trajectories simulated in the
dynamical transitions be long enough that they reach configurations
almost independent of their starting configurations. This avoids the
slow exploration, at the rate of a random walk, that would result if
the direction of motion were frequently randomized by stochastic
transitions.

In practice, the dynamics must be simulated with some finite step
size. The ``leapfrog'' method is generally used, with the following
steps being iterated some predefined number of times, $L$, with some
specified step size, $\epsilon$:
\begin{eqnarray}
  \pvec(\tau+{\textstyle{\epsilon \over 2}}) & = &
    \pvec(\tau)
     \ -\ {\textstyle{\epsilon \over 2}}\, \nabla E(\qvec(\tau))
\label{leaps}\\[3pt]
  \qvec(\tau+\epsilon) & = &
    \qvec(\tau)
     \ +\ \epsilon\, \pvec(\tau+{\textstyle{\epsilon \over 2}})
\\[3pt]
  \pvec(\tau+\epsilon) & = &
    \pvec(\tau+{\textstyle{\epsilon \over 2}})
     \ -\ {\textstyle{\epsilon \over 2}}\, \nabla E(\qvec(\tau+\epsilon))
\label{leape}\end{eqnarray}
This discretized mapping still preserves phase space volume exactly.
However, with a finite $\epsilon$, it does not leave $H$ exactly
constant. This will introduce some systematic error into the sampling.

\section{The hybrid Monte Carlo algorithm}

The systematic error of the dynamical method is eliminated in the
hybrid Monte Carlo algorithm \cite{duane} by considering the end-point
of the trajectory found with the leapfrog method to be merely a
candidate for the next state of the Markov chain, to be accepted or
rejected as in the Metropolis Monte Carlo algorithm \cite{metropolis}.

Acceptance or rejection of the candidate state is based on the amount,
$\Delta H$, by which $H$ for the candidate state exceeds $H$ for the current
state. The probability of acceptance, $a(\Delta H)$, is given by
\begin{eqnarray}
  a(\Delta H) & = & \min(1,\exp(-\Delta H))
\label{acc}\end{eqnarray}
Thus candidate states with lower $H$ are always accepted, while those
with higher $H$ are accepted with probability $\exp(-\Delta H)$. If
the candidate state is rejected, the new state is the same as the
current state (and is counted again in the average of equation (\ref{exp})).

The validity of this procedure for producing a sample from the
Boltzmann distribution is more easily seen if we imagine that the
trajectory for a dynamical transition is computed using a value for
$\epsilon$ whose sign is chosen at random, with positive and negative
values being equally likely. The leapfrog method (equations
(\ref{leaps}) to (\ref{leape})) is time-reversible, so that a
``forward'' trajectory, with a positive $\epsilon$, and a ``backward''
trajectory, with the corresponding negative $\epsilon$, are inverses
of each other, a fact crucial to the justification of the algorithm.
In fact, usual practice is to always use a positive $\epsilon$, since
an effect equivalent to randomly choosing its sign is produced in any
case by the randomization of the direction of $\pvec$ in the
stochastic transitions.

To show that the Boltzmann distribution (equation (\ref{distqp})) is
invariant under such dynamical transitions, it suffices to show that
these transitions satisfy the condition known as ``detailed balance''
--- that the probability of a transition from $A$ to $B$ occurring is
the same as that for a transition from $B$ to $A$, given that the
start state is Boltzmann distributed.

To see that detailed balance holds, consider a small region of volume
$\delta V$ around the point $A = (\qvec_A,\pvec_A)$. Suppose that a
forward trajectory from $A$ leads to the point $B = (\qvec_B,\pvec_B)$.
The other points in the region around $A$ will lead to a region around
$B$, which will also have volume $\delta V$, since the leapfrog method
preserves phase space volume. Due to time reversibility, a backward
trajectory from $B$ will lead to $A$. The detailed balance condition
with respect to the regions around $A$ and $B$ can now be written as
\begin{eqnarray}
\lefteqn{p(\qvec_A,\pvec_A)\, \delta V \cdot {\textstyle {1\over2}}
    \cdot a(H(\qvec_B,\pvec_B)\!-\!H(\qvec_A,\pvec_A))} \rule{20pt}{0pt}
\nonumber \\[3pt]
& = &
  p(\qvec_B,\pvec_B)\, \delta V \cdot {\textstyle {1\over2}}
    \cdot a(H(\qvec_A,\pvec_A)\!-\!H(\qvec_B,\pvec_B))
\end{eqnarray}
The left side of the above equation is the probability of moving from
the region around $A$ to the region around $B$. The first factor is
the Boltzmann probability for being in the region around $A$ at the
start, the second factor (${1\over2}$) is the probability of selecting
a forward direction for the trajectory, and the third factor is the
probability that this trajectory will be accepted. The right side
expresses the probability of moving from the region around $B$ to the
region around $A$ in analogous fashion. The equality of these two
probabilities is seen by substituting from equations (\ref{distqp})
and (\ref{acc}). The detailed balance condition can be similarly
verified for the case where a backward trajectory from $A$ is chosen.

The Boltzmann distribution is also invariant with respect to the
stochastic transitions, which simply replace $\pvec$ with a value
chosen from its Boltzmann distribution. Thus, if (as we expect) the
Markov chain is ergodic, it will have the Boltzmann distribution as
its unique stationary distribution.

\section{An acceptance procedure using windows}

The standard hybrid Monte Carlo algorithm can be generalized to
consider a window of states at the end of the trajectory as candidate
destinations for a dynamical transition, rather than just a single
end state. In order to maintain detailed balance, a possible move to
this window must be considered in relation to an equal-sized window
around the current state, with the location of the current state
within that window being determined randomly. Due to this later
requirement, the trajectory may have to be computed for some number of
steps in the reverse of its primary direction. Acceptance or rejection
of a move between windows is based on the total probability of the
states they contain. Whichever window is selected, a particular state
within that window is then picked according to the Boltzmann
probabilities.

This procedure can significantly increase the probability of accepting
a move, as will be discussed in Section 6. First, though, I will
define the algorithm in more detail.

To begin, values for the total number of steps in the trajectory, $L$,
the base step size, $\epsilon_0$, and the window size, $W$, are
selected from some fixed distribution, with $1 \leq W \leq L+1$. A
forward or backward direction, $\lambda$, for the trajectory is then
chosen, with equal probabilities for $\lambda=+1$ and
$\lambda=-1$, and an offset, $K$, for the current state within the
start window is selected, with $K \in \{0,\ldots,W\!-1\}$, each
possible value being equally likely.

A reverse portion of the trajectory is then computed by applying the
leapfrog method with a step size of $\epsilon = -\lambda\epsilon_0$,
beginning with the start state, $X(0)$. This is done for $K$
iterations, producing states labeled $X(-1), \ldots, X(-K)$. The
start state is then restored, and the leapfrog method is applied with
a step size of $\epsilon = +\lambda\epsilon_0$ for $L-K$ iterations,
producing states $X(1),\ldots,X(L-K)$.

The ``reject'' window at the front of the trajectory, around the
current state, is defined as the set $\calR = \{X(-K),\ldots,X(-K+W-1)\}$.
The ``accept'' window at the far end of the trajectory is defined as
$\calA = \{X(L-K-W+1),\ldots,X(L-K)\}$. These windows may overlap.

The ``free energy'' for a window is defined as follows:
\begin{eqnarray}
  F(\calW) & = &
    - \log \left(\,\sum_{X\in\calW}\!\exp(-H(X))\right)
\end{eqnarray}
The free energies are used to decide whether the next state will come
from the accept window or the reject window. Using the acceptance
function of equation~(\ref{acc}), the accept window is selected
with probability $a(\Delta F)$ where $\Delta F = F(\calA) - F(\calR)$;
otherwise we remain in the reject window (which contains the current
state). Having decided on the window $\calW$, a particular state
within that window is then selected according to the probabilities
\begin{eqnarray}
  P(X) & = & \exp(-H(X)+F(\calW))
\end{eqnarray}
The state selected becomes the next state in the Markov chain.

When implementing the generalized algorithm, it is not necessary to
save all the states in the accept and reject windows. One need only
save the start state, so it can be restored after the reversed portion
of the trajectory has been calculated, along with a single state from
the accept window and a single state from the reject window, one or
the other of which will become the next state of the Markov chain.

In detail, let $F_{\calA}^i$ be the free energy for the first $i$
states that have been visited in the accept window, and let
$C_{\calA}^i$ be a state chosen according to the Boltzmann
probabilities from among these first $i$ states. These variables can be
calculated incrementally as new states in the accept window are
visited. To start, $F_{\calA}^0=\infty$ and $C_{\calA}^0$ is
undefined. When we visit the $i$-th state in the accept window,
$(\qvec_i,\pvec_i)$, we can calculate
\begin{eqnarray}
  F_{\calA}^i & = &
   -\log\left(\,\exp(-H(\qvec_i,\pvec_i))+\exp(-F_{\calA}^{i-1}) \right)
\\[6pt]
  C_{\calA}^i & = & \left\{\begin{array}{ll}
     C_{\calA}^{i-1}    &
       \mbox{with probability $\exp(-F_{\calA}^{i-1}+F_{\calA}^i)$} \\[3pt]
     (\qvec_i,\pvec_i) &
       \mbox{with probability $\exp(-H(\qvec_i,\pvec_i)+F_{\calA}^i)$}
  \end{array}\right.
\end{eqnarray}
Once all states have been visited, we will have $F(\calA) =
F_{\calA}^W$, and $C_{\calA}^W$ will be a state picked from $\calA$
according to the Boltzmann probabilities. Analogous variables,
$F_{\calR}^i$ and $C_{\calR}^i$ are maintained for the reject window.
Once all states have been seen, a decision as to whether to use the
accept window or the reject window can easily be made, and, in either
case, a state selected from the chosen window is available.

When $W\!=\!1$, the generalized algorithm is equivalent to the standard
hybrid Monte Carlo algorithm. When $W\!=\!L+1$ the procedure reduces to
simply picking a state from those anywhere along the trajectory in
accordance with their Boltzmann probabilities. Some simplification in
the implementation is then possible.

\section{Validity of the generalized algorithm}

To demonstrate the validity of the generalized algorithm, we need to
show that the detailed balance condition holds for the dynamical
transitions. As $L$, $W$, and $\epsilon_0$ are chosen from a fixed
distribution, independently of the current state, we may choose to
regard them as fixed, since if detailed balance holds for transitions
with any values of these parameters, it will hold for a mixture of
such transitions. It will prove necessary to average over the values
selected for $\lambda$ and $K$, however.

Detailed balance will be proved separately for transitions to a state
in the accept window and for those to a state in the reject window. If
the two windows overlap, as they will if $W>(L+1)/2$, there will for some
pairs of states be the possibility of transitions of either type. If
detailed balance holds for the two types of transitions individually,
however, it will also hold for the combination.

To prove detailed balance for transitions within the reject window,
note first that the set of possible trajectories (for the various
values of $\lambda$ and $K$) that start at state $A$ and that
include state $B$ in the reject window is the same as the set of
trajectories that start at $B$ and include $A$ in the reject window.
Here, a ``trajectory'' is defined as the set of states visited, along
with the subsets of states that make up the accept and reject windows.
In detail, if for the trajectory starting at $A=X_A(0)$, with
direction $\lambda_A$ and window offset $K_A$, we have $B=X_A(J)$
within the reject window, then the identical trajectory will be
produced by starting at $B=X_B(0)$, with direction
$\lambda_B=\lambda_A$ and window offset $K_B = K_A+J$, and
$A=X_B(-J)$ will be in the reject window.

We can thus prove detailed balance separately for each such
trajectory. The probability of being in a small region of volume
$\delta V$ around $A$, of then picking values for $\lambda$ and $K$
that generate a particular trajectory for which a state in the region
of $B$ is in the reject window, of then chosing to pick a state from
the reject window, and of finally chosing the state in the region of
$B$ as the next state, is as follows:
\begin{eqnarray}
  p(\qvec_A,\pvec_A)\,\delta V \cdot {1 \over 2} \cdot {1 \over W}
   \cdot (1-a(F(\calA)\!-\!F(\calR))) \cdot \exp(-H(\qvec_B,\pvec_B)+F(\calR))
\end{eqnarray}
The probability of generating the same trajectory starting from the
region of $B$, and of then ending up in the region of $A$, is
\begin{eqnarray}
  p(\qvec_B,\pvec_B)\,\delta V \cdot {1 \over 2} \cdot {1 \over W}
   \cdot (1-a(F(\calA)\!-\!F(\calR))) \cdot \exp(-H(\qvec_A,\pvec_A)+F(\calR))
\end{eqnarray}
These are readily seen to be equal upon substituting from equation
(\ref{distqp}).

Detailed balance for transitions to a state in the accept window
follows similarly, using the fact that the set of possible
trajectories that start at state $A$ and that include state $B$ in the
accept window is the same as the set of trajectories that start at $B$
and include $A$ in the accept window, except that in the later
trajectories the accept and reject windows are exchanged. In detail,
if for the trajectory starting at $A=X_A(0)$, with window offset $K_A$
and direction $\lambda_A$, we have $B=X_A(J)$ within the accept
window, $\calA_A$, while $A$ is in the reject window, $\calR_A$, then
the trajectory produced by starting at $B=X_B(0)$, with window offset
$K_B = L-K_A-J$ and $\lambda_B = -\lambda_A$, will lead to
$A=X_B(J)$ being in the accept window, $\calA_B$, while $B$ is in the
reject window, $\calR_B$. Furthermore, we will have $\calA_A=\calR_B$
and $\calR_A=\calA_B$.

The probability of being in a small region of volume $\delta V$ around
$A$, of then picking values for $\lambda$ and $K$ that generate a
particular trajectory for which a state in the region of $B$ is in the
accept window, of then chosing to pick a state from the accept window,
and of finally chosing the state in the region of $B$ as the next
state, is as follows:
\begin{eqnarray}
  p(\qvec_A,\pvec_A)\,\delta V \cdot {1 \over 2} \cdot {1 \over W}
   \cdot a(F(\calA_A)\!-\!F(\calR_A)) \cdot
\exp(-H(\qvec_B,\pvec_B)+F(\calA_A))
\end{eqnarray}
For the probability of generating the same trajectory, but with accept and
reject windows exchanged, starting from the region of $B$, and of then
picking a state in the region of $A$, we have
\begin{eqnarray}
  p(\qvec_B,\pvec_B)\,\delta V \cdot {1 \over 2} \cdot {1 \over W}
  \cdot a(F(\calA_B)\!-\!F(\calR_B)) \cdot \exp(-H(\qvec_A,\pvec_A)+F(\calA_B))
\end{eqnarray}
Again, these are seen to be equal upon substituting from equations
(\ref{distqp}) and (\ref{acc}), remembering that $\calA_A=\calR_B$ and
$\calR_A=\calA_B$.

\section{Acceptance probability for the generalized algorithm}

Two situations where the use of windows will increase the acceptance
probability of the hybrid Monte Carlo algorithm are illustrated in
Figure 1.

In the trajectory of Figure 1(a), the energies of the states in the
reject window, at the start of the trajectory, are all approximately
equal to the energy of the current state, $H_0$. Most of the states in
the accept window, at the end of the trajectory, have energies in
the vicinity of $H_+$, much greater than that of the current state.
However, one state has a much lower energy, $H_-$.

If the standard hybrid Monte Carlo algorithm is applied in this
situation, with the end-point of the trajectory randomly picked
from the last $W$ states shown, the probability of acceptance will be
approximately $1/W$, since moves to states with energy $H_+$ would
almost certainly be rejected, while a move to the one state with
energy $H_-$ would be accepted.

In contrast, if the generalized algorithm is applied with a window of
size $W$, the move will always be accepted. The free energy of the
reject window will be
\begin{eqnarray}
  F(\calR) \ \approx\ -\log(W\exp(-H_0)) \ \approx\ H_0 - \log(W)
\end{eqnarray}
Assuming $H_+ \gg H_-$, the free energy of the accept window will be
\begin{eqnarray}
  F(\calA) \ \approx\ -\log((W-1)\exp(-H_+)+\exp(-H_-))
           \ \approx \ H_-
\end{eqnarray}
Thus, if $H_0 - H_- > \log(W)$, the move will be accepted, and the
particular state chosen within the accept window will almost certainly
be the one with energy $H_-$. Note that it is essential to this
example that a state in the accept window have lower energy than those
in the reject window. If this is not the case, the acceptance
probability is the same as for the standard algorithm.

The trajectory of Figure 1(b) illustrates a different, perhaps more
typical, situation where the use of windows also increases the
acceptance probability. Here, both windows contain approximately equal
numbers of low and high energy states. Note that the current state is
likely to be one of the low energy ones, since it is Boltzmann
distributed.

With the standard algorithm, the end state might equally well be one
of high energy or one of low energy. In the former case, the move
would likely be rejected, while in the later, it would have a good
probability of being accepted. The total acceptance probability will
thus be around $1/2$.

If the generalized algorithm is used with a window size large compared
to the time scale of energy fluctuations, however, the free energy of
both the accept and reject windows will be approximately equal, and
the acceptance probability will be near one. The particular state
chosen within the accept window will likely be one with low energy.

These examples show that the use of windows can be beneficial, but
they do not indicate the magnitude of the benefit, nor whether there
is any improvement in the asymptotic form of the time requirements as
system size increases. Indeed, for fixed values of $L$, $\epsilon_0$,
and $W$, the acceptance probability declines exponentially with
increasing system size, as for the standard algorithm. This scaling
behaviour is due to the free energies of the windows, and hence their
difference, being extensive quantities that increase in proportion to
system size.

\section{Performance for a system of uncoupled oscillators}

To gain some insight into the degree of benefit from using the
generalized algorithm, and into its scaling behaviour, I have tested
it empirically on systems of uncoupled oscillators. These simple
systems are meant to model more complex systems in which the
components are not completely independent, but interact only weakly.
The behaviour of the standard hybrid Monte Carlo algorithm for systems
of uncoupled oscillators has been analysed in detail by Kennedy and
Pendleton \cite{kennedy}.

The potential energy function for such a system is
\begin{eqnarray}
   E(\qvec) & = & {1 \over 2}\, \sum_{i=1}^{N}\, \omega_i^2 q_i^2
\end{eqnarray}
In the Boltzmann distribution with respect to $E$, each $q_i$ is
independent, and distributed as a Gaussian with zero mean and standard
deviation $1/\omega_i$. Since the operation of the hybrid Monte Carlo
algorithm is invariant with respect to translations and rotations of
the coordinates, these systems in fact model the behaviour of the
algorithm as applied to any multivariate Gaussian distribution.

Note that for the leapfrog method to be stable, with the error in $H$
remaining bounded even for long trajectories, it is necessary for the
step size to be less than $2/\omega_{\max}$, where $\omega_{\max}$ is
the largest of the $\omega_i$. For efficient exploration of the other
coordinates, the average length of a trajectory in fictitious time,
$T_t$, should be in the vicinity of $1/\omega_{\min}$.

In using these systems as a test bed, we must be clear concerning
which aspects of the system we expect to carry over to realistic
problems, and which will not. I assume here that the approximate
magnitudes of $\omega_{\min}$ and $\omega_{\max}$ and the general
distribution of the $\omega_i$ are known, and that these may be used
to select appropriate values for the step size and trajectory length.
I assume that their {\em exact\/} values would not be known in a
realistic system --- indeed, they will not be precisely defined, since
the components of the real system will not be completely uncoupled. To
model this, the step size (and hence the trajectory length as well)
was varied slightly at random, in order to avoid results that depend
on precise tuning of these parameters.

I will also assume that we are interested in systems for which the
ratio $\omega_{\max}/\omega_{\min}$ is large. This ratio is a measure
of the inherent difficulty of the problem, being (roughly) the number
of leapfrog steps required to generate an independent configuration.
In the experiments, the length in fictitious time of a trajectory was
chosen to be large enough for the acceptance probability to have
reached equilibrium. In a real system, however, the appropriate
trajectory length ($T_t \approx 1/\omega_{\min}$) might well be much
longer than this.

Since chosing any particular value for $T_t$ would be arbitrary, I
have for the most part evaluated the algorithms on the assumption that
$T_t$ is very large. The ``cost'' of a particular combination of
window size and average step size was accordingly taken to be
\begin{eqnarray}
  C & = & 1\,/\,(\oeps(1\!-\!\rho))
\label{cost}\end{eqnarray}
where $\rho$ is the probability of a move being rejected, and $\oeps$
is the average step size (recall that the actual step sizes used will
be slight perturbations around this average). This cost measure is
proportional to the number of energy gradient evaluations needed to
generate a given number of accepted moves of average length $T_t$,
provided $T_t$ is much greater than $T_w$, the length in fictitious
time of the windows. Note that this cost measure places no value on
transitions within the reject window, though these presumably improve
sampling at least somewhat.

If $T_w$ is comparable to $T_t$, then the above measure would not be
appropriate, since it ignores the effort expended in computing those
portions of the trajectory that lie before the current state and after
the new state. An appropriate cost in this case would be
$(1+T_w/T_t)\,/\,(\oeps(1\!-\!\rho))$.

I assume that for a given window length, $T_w$, and a given system
size, $N$, the value of $\oeps$ that minimizes $C$ can feasibly be
found, and that this minimal value of $C$ is thus the appropriate
measure of the cost of the algorithm for a particular window length.
If the rejection probability has the functional form
\begin{eqnarray}
  \rho & = & F(g(N)\oeps^{\,k})
\label{aform}\end{eqnarray}
then one can show by straightforward means that this minimum is
reached at a value of $\rho$ that is independent of $N$. The scaling
properties of the algorithm can then be found by determining how
$\oeps$ must decrease as $N$ increases in order to keep the rejection
probability constant. Note that we are measuring cost as the number of
evaluations of the energy gradient, not the time required for these
evaluations. Computation time per evaluation will vary with system
size in a manner which depends on the application, but which will
generally add an additional factor of at least $N$ to the total
computation time as a function of system size.

I have tested the standard and the generalized algorithms on systems
for which $N = 100$, $200$, $400$, $800$, $1600$, and $3200$. In each
case, the $\omega_i$ were selected randomly from the range $500$ to
$1000$, with a uniform distribution for $\log(\omega)$. Though I
assume that components with much smaller $\omega$ would also be
present in a real system, such were not actually included in the
simulation, in order to save time. Their inclusion would have had a
negligible effect on the acceptance probability, since the accuracy of
the leapfrog method is very high when $1/\omega$ is much greater than
the step size.

An average trajectory length of $T_t = 1 \approx 1000 \times
(1/\omega_{\max})$ was used. Several of the simulations were done with
$T_t = 2$ as well, and the results confirmed that the acceptance
probability had indeed reached equilibrium at $T_t = 1$.

For a given average step size, $\oeps$, and for a given window size,
$W$, the number of steps in the entire trajectory, $L$, was set so
that $T_t = \oeps(L\!-\!W\!+\!1)$. (This calculation accounts for the
average number of steps before and after the current and new states.)
The trajectory was then computed with these values of $L$ and $W$ and
with a value of $\epsilon_0$ randomly selected from the region within
1\% of the given $\oeps$.

Runs were done of the standard algorithm, for which $W=1$, and of the
generalized algorithm with a window length of $T_w = 0.20$, for which
the number of states in the window was $W = T_w/\oeps$. Some runs with
$T_w=0.05$ and $T_w=0.10$ were done as well, with results that were
similar to but not quite as good as those for $T_w=0.20$.  Values for
$\oeps$ of $\ldots,\, 0.000707,\, 0.000841,\, 0.001000,\, 0.001189,\,
\ldots$, in geometric steps of $2^{1/4}$ were used. In each run, 1000
trajectories were generated starting from position and momentum
coordinates picked from the Boltzmann distribution, independently for
each trajectory. The proportion of rejected moves, $\hat\rho$, was
taken to be an estimate of the rejection probability, $\rho$. The
standard error for this estimate is $\pm 0.016$ at $\rho=0.5$,
declining to $\pm 0.009$ at $\rho=0.1$ or $\rho=0.9$. An estimate,
$\widehat C$, for the cost follows from equation (\ref{cost}).

The results are shown in Figure 2, which plots the estimated rejection
probability and consequent cost for each value of $N$ and for various
values of $\oeps$, for both the standard algorithm and the generalized
algorithm with $T_w=0.20$. When the best value of $\oeps$ is used in
each case, the cost of the generalized algorithm is roughly half that
of the standard algorithm, with some indication that the advantage of
the generalized algorithm may be greater for the larger $N$.
Interestingly, the rejection probability with the optimal $\oeps$ is
significantly lower for the generalized algorithm than for standard
algorithm.

In analysing this data further, we can first compare with the analytic
results of \cite{kennedy}. They derive the following expression for
the rejection probability of the standard algorithm applied to a
system of uncoupled oscillators (adapted from their equation (2.10)):
\begin{eqnarray}
  \rho \ \approx \ \erf \left(\sqrt{\textstyle N\epsilon^4\sigma/32} \right)
\end{eqnarray}
where $\sigma = N^{-1} \sum_i \omega_i^4 \sin^2(\omega_i T_t) / 4$.
Assuming that the small random variation in $\epsilon$, and hence
$T_t$, is enough to randomize the phases in this sum, and using the
fact that ${1\over2\pi}\int_0^{2\pi} \sin^2(x) dx = {1\over2}$, we get
that, for the experiment described here, the rejection probability
should be
\begin{eqnarray}
  \rho \ \approx \ \erf\left(\sqrt{\textstyle N\oeps^{\,4} \osig/256}\right)
\label{stpred}\end{eqnarray}
where $\osig = N^{-1} \sum_i \omega_i^4 \approx 3.38\times10^{11}$.
The rejection probability thus has the form of equation (\ref{aform}).
As system size increases, $\oeps$ should be scaled as $N^{-1/4}$ in
order to keep the rejection probability constant, and the cost will
grow as $N^{1/4}$.

In Figure 3, $\hat\rho$ is plotted against $N\oeps^{\,4}$ for all runs
of the standard algorithm. As expected, $\rho$ appears to be a
function of $N\oeps^{\,4}$, as the spread in the data points in
comparable to the standard error. (Note that here, and in Figure 4,
the standard error is apparently larger for smaller values of $\rho$,
due to the logarithmic scale.) Comparison with the exact predictions
of equation (\ref{stpred}) (not shown in the figure) shows a good fit,
except for a slight departure for large values of the rejection rate.

We can hypothesize that the rejection probability of the generalized
algorithm for a given window length will also be a function of
$N\oeps^{\,4}$. If this is the case, the cost for the generalized
algorithm will also scale as $N^{1/4}$, though there could, as seen
above, be an improvement in the constant factor over the standard
algorithm. To test this hypothesis, Figure 3 also shows $\hat\rho$
plotted against $N\oeps^{\,4}$ for all runs of the generalized algorithm
with $T_w=0.20$. A tendency of the curves for the larger $N$ to lie
below those for the smaller $N$ is apparent, leading one to reject
this hypothesis.

The data for the generalized algorithm is better fit by the
alternative hypothesis that the rejection probability is a function of
$N^{7/8}\oeps^{\,4}$. This is seen in Figure 4. For comparison, the
data for the standard algorithm is also plotted under this assumption,
and it is clear that in this case the fit is bad.

If this hypothesis is true, then the step size for the generalized
algorithm should be scaled as $N^{-7/32}$ to maintain a constant
rejection rate, and the cost will consequently grow with system size
as $N^{7/32}$, an improvement over the $N^{1/4}$ scaling of the
standard algorithm. This result must be regarded as tentative,
however. It seems possible that for very large $N$ the window length,
$T_w$, might have to increase at some rate in order to maintain good
scaling behaviour. This would ultimately affect the cost, once $T_w$
became comparable to $T_t$. Behaviour could also conceivably depend on
the exact distribution of the $\omega_i$. A theoretical analysis is
thus needed to gain a better understanding of the performance of the
generalized algorithm. It is clear, however, that at the very least,
it can improve performance by a significant constant factor.

\section{Variations on the algorithm}

Two variations on the generalized algorithm described here are worth
mentioning.

First, when a move to the accept window is rejected, it is valid to
simply remain at the current state, rather than selecting a new state
from the reject window according to the Boltzmann probabilities. This
follows from the fact that detailed balance was shown above to hold
independently for the accept and reject transitions. Detailed balance
thus continues to hold if all reject transitions are eliminated.

With this variation, the overhead of saving states is somewhat
reduced. Typically, however, this overhead is small compared to the
cost of evaluating the gradient of the energy, and its elimination may
not be worth giving up the additional exploration provided by the
reject transitions. Note that it is in any case necessary to visit all
the states in the reject window, in order to compute its free energy.

A second variation may be useful when the ideal step size is not known
{\em a priori}. In such cases, we may have to use step sizes selected
at random from a fairly broad distribution. Trajectories computed with
a step size that is too large will result in large changes in $H$ and
are unlikely to be accepted.

To save computation, we can terminate such trajectories early,
stopping whenever a single leapfrog step changes $H$ by an amount,
positive or negative, whose magnitude is greater than some threshold.
The state reached after this large change in $H$ is not included. Such
truncated trajectories will have smaller than normal accept and/or
reject windows, but examination of the proof of validity in Section~5
shows that they may validly be treated the same way as normal
trajectories. The accept window for a truncated trajectory may, in
fact, be null, in which case the move is rejected. (The reject window
will always contain at least the current state.)

This variation is applicable to the standard hybrid Monte Carlo
algorithm, where the window size is one. In this case, all truncated
trajectories are rejected, with the current state remaining unchanged.

\section*{Acknowledgements}

I thank David MacKay for helpful comments. This work was supported by
the Natural Sciences and Engineering Research Council of Canada, and
by the Ontario Information Technology Research Centre.

\pagebreak

\pagebreak

\begin{figure}[p]
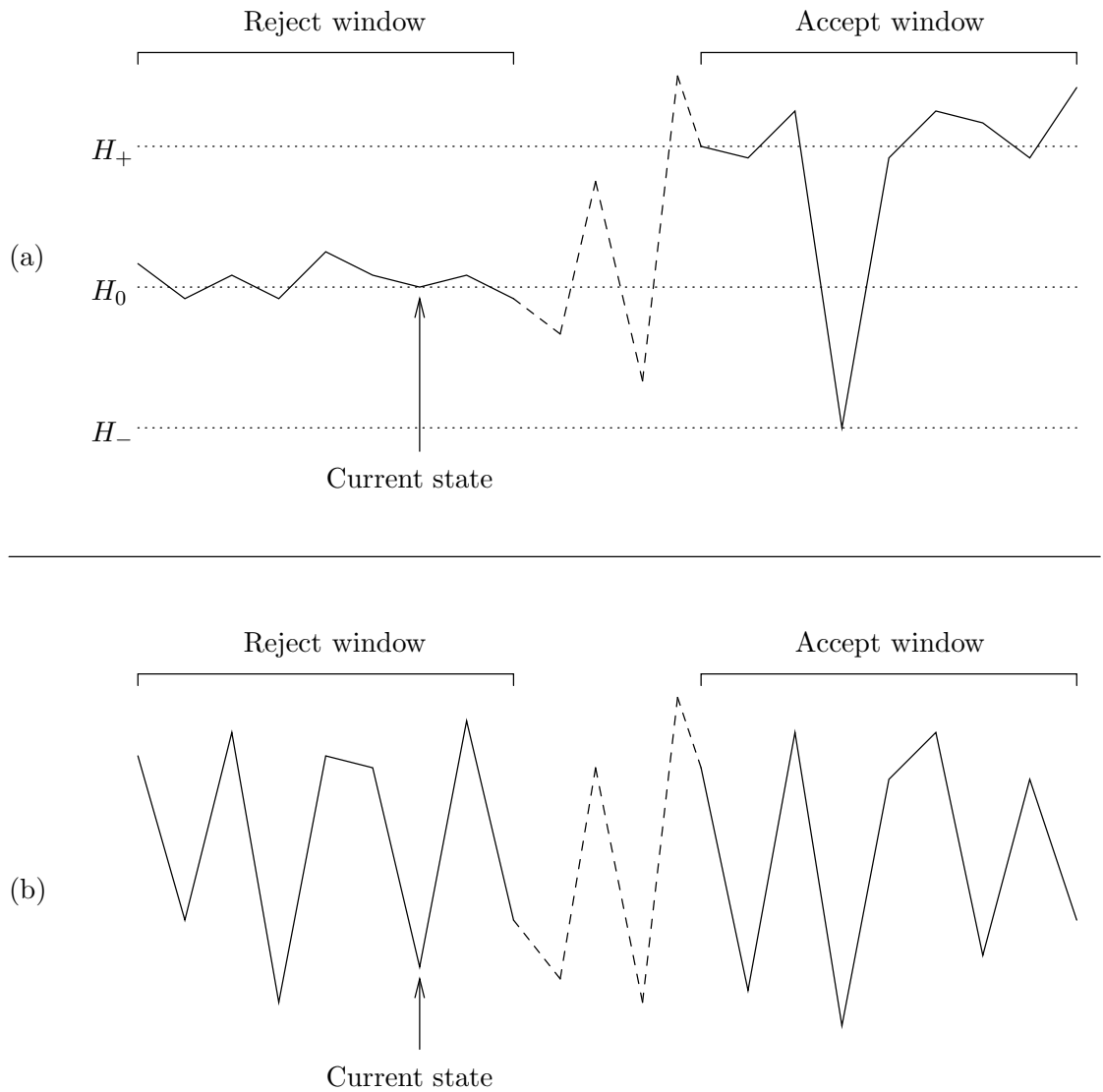


\catcode`@=11
\expandafter\ifx\csname graph\endcsname\relax \alloc@4\box\chardef\insc@unt
\graph\fi
\expandafter\ifx\csname graphtemp\endcsname\relax \newdimen\graphtemp\fi
\catcode`@=12
\setbox\graph=\vtop{
  \vbox to0pt{\hbox{
    \special{pn 8}%
    \special{pa 0 2843}%
    \special{pa 5811 2843}%
    \special{fp}%
    \special{pa 688 2155}%
    \special{pa 5686 2155}%
    \special{dt 0.04}%
    \special{pa 688 1281}%
    \special{pa 938 1468}%
    \special{pa 1188 1343}%
    \special{pa 1438 1468}%
    \special{pa 1688 1218}%
    \special{pa 1937 1343}%
    \special{pa 2187 1406}%
    \special{pa 2437 1343}%
    \special{pa 2687 1468}%
    \special{fp}%
    \special{pa 5686 343}%
    \special{pa 5436 718}%
    \special{pa 5186 531}%
    \special{pa 4936 468}%
    \special{pa 4686 718}%
    \special{pa 4436 2155}%
    \special{pa 4186 468}%
    \special{pa 3936 718}%
    \special{pa 3686 656}%
    \special{fp}%
    \special{pa 688 656}%
    \special{pa 5686 656}%
    \special{dt 0.04}%
    \special{pa 688 1406}%
    \special{pa 5686 1406}%
    \special{dt 0.04}%
    \special{pa 2687 1468}%
    \special{pa 2936 1656}%
    \special{da 0.05}%
    \special{pa 2936 1656}%
    \special{pa 3124 843}%
    \special{da 0.05}%
    \special{pa 3124 843}%
    \special{pa 3374 1905}%
    \special{da 0.05}%
    \special{pa 3374 1905}%
    \special{pa 3561 281}%
    \special{da 0.05}%
    \special{pa 3561 281}%
    \special{pa 3686 656}%
    \special{da 0.05}%
    \special{pa 3686 218}%
    \special{pa 3686 156}%
    \special{pa 5686 156}%
    \special{pa 5686 218}%
    \special{fp}%
    \special{pa 688 218}%
    \special{pa 688 156}%
    \special{pa 2687 156}%
    \special{pa 2687 218}%
    \special{fp}%
    \special{pa 2187 2280}%
    \special{pa 2187 1468}%
    \special{fp}%
    \special{pa 2212 1568}%
    \special{pa 2187 1468}%
    \special{pa 2162 1568}%
    \special{fp}%
    \special{pa 688 3904}%
    \special{pa 938 4779}%
    \special{pa 1188 3779}%
    \special{pa 1438 5217}%
    \special{pa 1688 3904}%
    \special{pa 1937 3967}%
    \special{pa 2187 5029}%
    \special{pa 2437 3717}%
    \special{pa 2687 4779}%
    \special{fp}%
    \special{pa 5686 4779}%
    \special{pa 5436 4029}%
    \special{pa 5186 4967}%
    \special{pa 4936 3779}%
    \special{pa 4686 4029}%
    \special{pa 4436 5342}%
    \special{pa 4186 3779}%
    \special{pa 3936 5154}%
    \special{pa 3686 3967}%
    \special{fp}%
    \special{pa 2687 4779}%
    \special{pa 2936 5092}%
    \special{da 0.05}%
    \special{pa 2936 5092}%
    \special{pa 3124 3967}%
    \special{da 0.05}%
    \special{pa 3124 3967}%
    \special{pa 3374 5217}%
    \special{da 0.05}%
    \special{pa 3374 5217}%
    \special{pa 3561 3592}%
    \special{da 0.05}%
    \special{pa 3561 3592}%
    \special{pa 3686 3967}%
    \special{da 0.05}%
    \special{pa 3686 3529}%
    \special{pa 3686 3467}%
    \special{pa 5686 3467}%
    \special{pa 5686 3529}%
    \special{fp}%
    \special{pa 688 3529}%
    \special{pa 688 3467}%
    \special{pa 2687 3467}%
    \special{pa 2687 3529}%
    \special{fp}%
    \special{pa 2187 5467}%
    \special{pa 2187 5092}%
    \special{fp}%
    \special{pa 2212 5192}%
    \special{pa 2187 5092}%
    \special{pa 2162 5192}%
    \special{fp}%
    \special{pn 11}%
    \graphtemp=.6ex \advance\graphtemp by 0.687in
    \rlap{\kern 0.438in\lower\graphtemp\hbox to 0pt{$H_+$\hss}}
    \graphtemp=.6ex \advance\graphtemp by 1.437in
    \rlap{\kern 0.438in\lower\graphtemp\hbox to 0pt{$H_0$\hss}}
    \graphtemp=.6ex \advance\graphtemp by 2.187in
    \rlap{\kern 0.438in\lower\graphtemp\hbox to 0pt{$H_-$\hss}}
    \graphtemp=.6ex \advance\graphtemp by 2.437in
    \rlap{\kern 1.688in\lower\graphtemp\hbox to 0pt{Current state\hss}}
    \graphtemp=.6ex
    \rlap{\kern 1.25in\lower\graphtemp\hbox to 0pt{Reject window\hss}}
    \graphtemp=.6ex
    \rlap{\kern 4.186in\lower\graphtemp\hbox to 0pt{Accept window\hss}}
    \graphtemp=.6ex \advance\graphtemp by 1.25in
    \rlap{\lower\graphtemp\hbox to 0pt{(a)\hss}}
    \graphtemp=.6ex \advance\graphtemp by 5.624in
    \rlap{\kern 1.688in\lower\graphtemp\hbox to 0pt{Current state\hss}}
    \graphtemp=.6ex \advance\graphtemp by 3.311in
    \rlap{\kern 1.25in\lower\graphtemp\hbox to 0pt{Reject window\hss}}
    \graphtemp=.6ex \advance\graphtemp by 3.311in
    \rlap{\kern 4.186in\lower\graphtemp\hbox to 0pt{Accept window\hss}}
    \graphtemp=.6ex \advance\graphtemp by 4.624in
    \rlap{\lower\graphtemp\hbox to 0pt{(b)\hss}}
    \kern 5.811in
  }\vss}%
  \kern 5.89in
}

\centerline{\box\graph}
\caption{Two situations where the use of windows increases the acceptance
         probability. The graphs show the change in total energy along
         two hypothetical trajectories. }
\end{figure}


\vspace*{0.5in}

\begin{thebibliography}{9}

\bibitem{anderson}
  H. C. Andersen, Molecular dynamics simulations at constant
  pressure and/or temperature, {\em Journal of Chemical Physics},
  {\bf 72}, 2384-2393 (1980).

\bibitem{duane}
  S. Duane, A. D. Kennedy, B. J. Pendleton and D. Roweth,
  Hybrid Monte Carlo, {\em Physics Letters B}, {\bf 195}, 216-222 (1987).

\bibitem{kennedy}
  A. D. Kennedy and B. Pendleton, Acceptances and autocorrelations
  in hybrid Monte Carlo, {\em Nuclear Physics B (Proc. Suppl.)}, {\bf 20},
  118-121 (1991).

\bibitem{metropolis}
  N. Metropolis, A. W. Rosenbluth, M. N. Rosenbluth, A. H. Teller
  and E. Teller, Equation of state calculations by fast computing
  machines, {\em Journal of Chemical Physics}, {\bf 21}, 1087-1092 (1953).

\bibitem{toussaint}
  D. Toussaint, Introduction to algorithms for Monte Carlo
  simulations and their application to QCD, {\em Computer Physics
  Communications}, {\bf 56}, 69-92 (1989).

\end{thebibliography}
\end{document}